\documentclass[11pt,letterpaper,reqno]{amsart}
\usepackage{xcolor}
\usepackage{graphicx}
\usepackage{subcaption}

 \theoremstyle{definition}
 
 \theoremstyle{remark}

 \numberwithin{equation}{section}

\usepackage{ amsmath, amsthm, amssymb, amsfonts}

\newcommand{\bR}{\mathbb{R}}

\calclayout

\begin{document}

\title{Harold Widom's work in random matrix theory}
\author{Ivan Z. Corwin}
\address{Columbia University\,\
New York, NY 10027, USA}
\email{corwin@math.columbia.edu}
\author{Percy A.~Deift }
\address{Courant Institute of Mathematical Sciences\,\
New York, NY 10003, USA}
\email{deift@courant.nyu.edu}
\author{Alexander R.~Its }
\address{Department of Mathematical Sciences\\
Indiana University Purdue University  Indianapolis\\
Indianapolis, IN 46202-3216, USA}
\email{aits@iu.edu}

\begin{abstract}
This is a survey of Harold Widom's work in random matrices. We start with his pioneering
papers on the sine-kernel determinant, continue with his and Craig Tracy's  groundbreaking results
concerning the distribution functions of random matrix theory, touch on the remarkable universality of the Tracy-Widom distributions in mathematics and physics, and close with Tracy and Widom's remarkable work on the asymmetric simple exclusion process.
\end{abstract}

\thanks{This work was supported in part by the National Science Foundation through grants DMS:1937254, DMS:1811143, DMS:1664650 of the first author,  DMS:1001886 of the second author
and DMS:1001777, DMS:1955265 of the third author. The first author is also supported through a Packard Fellowship in Science and Engineering and a W.M. Keck Foundation Science and Engineering Grant.}
\maketitle

\section{Introduction}
The distributions of random matrix theory govern  the
statistical properties of  a wide variety of  large systems which  do not  obey the usual
laws of classical probability. Such systems appear in
many different areas of applied science and technology, including
heavy nuclei, polymer growth, high-dimensional data  analysis, and
certain percolation processes. The four distribution functions play
a particularly important  role in the mathematical apparatus of random matrices. The first one describes the
``emptiness formation probability'' in the bulk of the spectrum of a large random matrix,
and it is explicitly given in terms of the  sine kernel Fredholm determinant.
The second, the third, and the forth  distributions  are given in terms of the Airy kernel
Fredholm determinant, and they describe the edge fluctuations
of the eigenvalues in the large size limit of the matrices taken from the three
classical Gaussian ensembles, unitary (GUE), orthogonal (GOE) and symplectic (GSE).
The last  three of these distributions are known now as the Tracy-Widom
distribution functions.

A key analytical observation concerning the
distribution functions of random matrix theory, which was made on many occasions in the papers
\cite{74,75,80,84,95,96}, is that they satisfy certain nonlinear integrable PDEs.
This property, which  generalizes the  first result of this type established in \cite{ JMMS}
for the sine kernel determinant,  follows in turn  from a remarkable Fredholm determinant
representation for the random matrix distributions. The  existence of such representations
in several important examples  beyond the sine kernel case  was also first shown in the above mentioned
papers.

What follows is an  overview of the principal contributions of Harold Widom to random matrix theory. We start
with his papers \cite{72} and \cite{83} on the sine kernel determinant, which brought  mathematical rigor to the
theory of this, the first universal distribution function of random matrix theory.


\section{The sine-kernel determinant}

Let $J = \cup_{k=0}^{n}J_k = \cup_{k=0}^{n}(a_k, b_k) $ be a union of $n+1$ disjoint intervals in $\bR$.
Consider the Fredholm determinant
$$
P_x= \det \left( 1 - K_{\textrm{sine}}\right),
$$
where $K_{\textrm{sine}}$ is the trace class operator in $L_2\Bigl(J; dz\Bigr)$  with kernel
$$
K_{\textrm{sine}}(z,z') = \frac{\sin x(z-z')}{\pi (z-z')}.
$$
The determinant $P_x$ plays a central role in random matrix theory. Indeed, it is the probability
of finding no eigenvalues in the union of intervals $\frac{xJ}{\pi} := \cup_{k=0}^{n}\left(\frac{xa_k}{\pi}, \frac{xb_k}{\pi}\right)$
for a random Hermitian matrix chosen from the Gaussian Unitary Ensemble (GUE), in the bulk scaling limit with mean
spacing 1 (see \cite{M} and Figure \ref{fig:sine}).  Moreover, in the one interval case,  the second derivative of $P_x$ describes the distribution
of normalized  spacings of eigenvalues of large random GUE matrices (see, e.g. \cite{Deift0}).
The determinant $P_x$ also appears in quantum and statistical mechanics. For instance, it  describes the emptiness formation probability
in the one-dimensional impenetrable Bose gas \cite{L} and  the gap probability in the one dimensional Coulomb
gas, at inverse temperature $\beta = 2$ (see, e.g. \cite{D1}).
The key analytical issue related to  $P_x$ is its large $x$ behavior, i.e., the ``large gap asymptotics''.
Harold Widom made major contributions to the resolution of this question.

\begin{figure}[h]
\centering
\scalebox{0.6}{\includegraphics{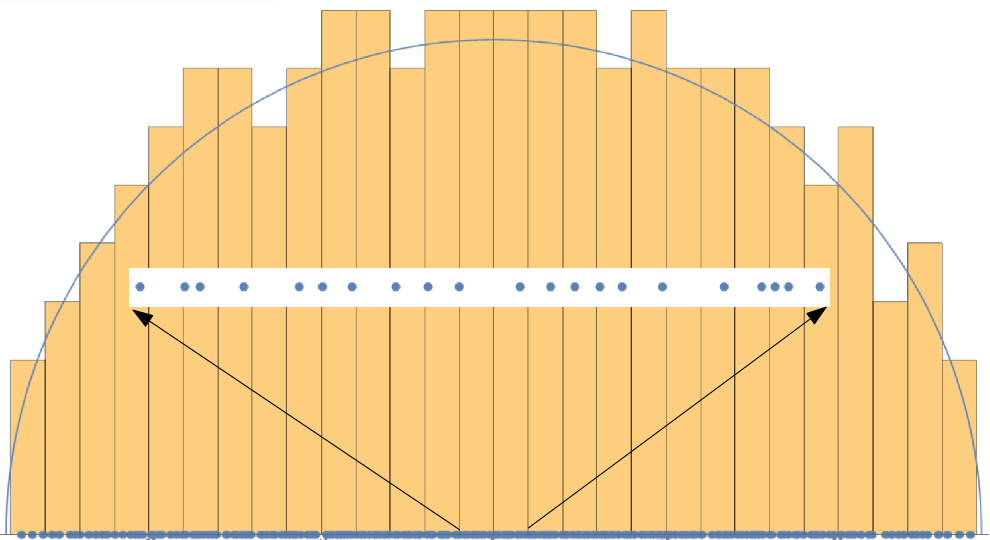}}
\captionsetup{width=.9\linewidth}
\caption{The eigenvalues of a large ($2000\times 2000$) GUE matrix. The full point process sits on the bottom of the figure, and the histogram records the density of points (approximating the Wigner semi-circle). Inlaid is a zoom-in of the point process. As the size of the GUE matrix goes to infinity, this converges to $\beta=2$ sine process whose gap probabilities are discussed here.}
\label{fig:sine}
\end{figure}

In the one interval case, $n =0$, after rescaling and translation, we may assume $J = (-1, 1)$.
For this case, in 1973, des Cloizeaux and Mehta \cite{dCM} showed that as $x\to\infty$,
\begin{equation}
\ln P_x=-{x^2\over 2} -{1\over 4}\ln x+ c_0 +o(1),\label{as1}
\end{equation}
for some constant $c_0$. In 1976, Dyson \cite{D2} showed that $P_x$ in fact has a
full asymptotic expansion of the form
\begin{equation}\label{dyson_as}
\ln P_x=-{x^2\over 2} -{1\over 4}\ln x+ c_0 +{a_1\over x}+ {a_2\over
x^2}+\cdots.
\end{equation}
Dyson identified all the constants $c_0$, $a_1$, $a_2$, $\dots$. Of
particular interest is the constant $c_0$, which he found to be
\begin{equation}
c_0={1\over 12}\ln 2+ 3\zeta'(-1),\label{c0}
\end{equation}
where $\zeta(z)$ is the Riemann zeta-function. It should be noted that Dyson obtained this result
using one of the early results of Widom \cite{33}  on the asymptotics
of Toeplitz determinants with symbols supported on circular arcs.

The results in \cite{dCM} and \cite{D2} were not fully rigorous. In \cite{72},
using an adaptation of Szeg\H{o}'s classical method to the continuous analogues
of orthogonal polynomials (the so-called Krein functions),  Widom gave the first
rigorous proof of the leading asymptotics in
(\ref{as1}) in the form
\begin{equation}
\ln P_x=-{x^2\over 2}(1+o(1))
\end{equation}
as $x \to \infty$. Actually, Widom proved a slightly stronger result,
$$
\frac{d}{dx}\ln P_x = -x + O(1)
$$
as $x \to \infty$.  In addition, Widom computed the leading asymptotics of  $\textrm{tr}\left( (1- K_{\textrm{sine}})^{-1}K_{\textrm{sine}}\right)$, which is the ratio
of the probability  that there is at most one eigenvalue in the interval $\left(\frac{-x}{\pi}, \frac{x}{\pi}\right)$
to $P_x$. In  the subsequent paper \cite{83}, Widom considered the multi-interval case, $n > 0$, and
showed  that as $x \to \infty$,
$$
\frac{d}{dx}\ln P_x = c_1 x + c_2(x) + o(1)
$$
where $c_1$ is a negative constant and $c_2(x)$ is a certain bounded oscillatory function
of $x$, which was described up to the solution of a Jacobi inversion problem{\footnote{ An
explicit formula for $c_2(x)$, in terms of the Riemann theta-function, was obtained  later in \cite{DIZ}.}}.
The method of \cite{83} is  a further development of the approach in \cite{72}. As in \cite{72}, Widom also computes the
leading asymptotics of  $\textrm{tr}\left( (1- K_{\textrm{sine}})^{-1}K_{\textrm{sine}}\right)$.

The formula (\ref{c0}) for  $c_0$ in \cite{D2} was in the form of conjecture: A
rigorous proof of (\ref{c0}) was only given in 2004. In fact, two proofs of  this formula
were given independently in \cite{K} and \cite{E}.
It is remarkable that the proof of (\ref{c0}) in \cite{K} again uses (in conjunction with the nonlinear steepest descent method)
Widom's computation in  \cite{33}.


\section{The Tracy-Widom distribution functions}
The  joint probability densities of  the eigenvalues of  random matrices
from the GOE, GUE and GSE ensembles 
are given by
\begin{equation}\label{eigenprob}
P_{N\beta}(\lambda_1, ..., \lambda_N) =
Z^{-1}_{N\beta}\,e^{-\frac{1}{2}\beta\sum_{j=1}^{N}\lambda^{2}_j}
\prod_{j<k}|\lambda_j - \lambda_k|^{\beta},
\end{equation}
where $Z_{N\beta}$ is a normalization constant (or  partition function) and
$$
\beta := \begin{cases}
1 & \textrm{for GOE},\cr
2 & \textrm{for GUE},\cr
4 & \textrm{for GSE}.
\end{cases}
$$
The famous Tracy-Widom distribution functions, commonly denoted
as $F_{\beta}(t)$,  describe the edge fluctuation
of the eigenvalues in the large $N$ limit. They are defined via the
scaling limits
\begin{equation}\label{Airy1}
F_{\beta}(t) = \lim_{N\to \infty}\textrm{Prob}\,\bigl((\lambda_{\max} -\sqrt{2N})2^{1/2}N^{1/6}
\leq t\bigl),
\end{equation}
where $\lambda_{\max}$ is the largest eigenvalue drawn from the ensembles with density (\ref{eigenprob}).
The central theme of the series of papers \cite{74,75,80,84,96} is the following analytical description of
these distributions, which links them to the theory of integrable systems.

Let $K_{\textrm{Airy}}$ be the  trace-class operator in $L_2\bigl((t, \infty); dz\bigr)$ with kernel
\begin{equation}\label{Airy0}
K_{\textrm{Airy}}(z, z') = \frac{\textrm{Ai}(z)\textrm{Ai}'(z') - \textrm{Ai}'(z)\textrm{Ai}(z')}{z - z'},
\end{equation}
where $\textrm{Ai}(z)$ is the Airy function,
$$
\textrm{Ai}(z) = \frac{1}{2\pi}\int_{-\infty}^{\infty}e^{-\frac{i}{3}s^3 - izs}ds.
$$
Then, the  $\beta =2$  Tracy-Widom distribution, $F_{2}(t)$, is given
by the {\it Airy-kernel}  Fredholm determinant (see \cite{For})
\begin{equation}\label{Airy2}
F_{2}(t) = \det \left(1 - K_{\textrm{Airy}}\right).
\end{equation}
Moreover, as  shown in \cite{75}, the following  formula is
valid for the  Fredholm determinant on the right hand side of (\ref{Airy2}):
\begin{equation}\label{TW1}
 \det(1 - K_{\textrm{Airy}}) = \exp\left\{ - \int_{t}^{\infty}(s-t)u^2(s)ds\right\},
\end{equation}
where $u(x)$ is the Hastings-McLeod solution of the {\it second Painlev\'e
equation}, i.e. the solution of the ODE
\begin{equation}\label{painleve}
u_{tt} = tu +2u^3,
\end{equation}
uniquely determined by the boundary condition,
\begin{equation}\label{painleve2}
u(t) = \frac{1}{2\sqrt{\pi}}t^{-1/4}e^{-\frac{2}{3}t^{3/2}}(1 + o(1)), \quad t \to +\infty.
\end{equation}
Similar Painlev\'e  representations for the other two Tracy-Widom distribution functions
have the form \cite{84}
\begin{equation}\label{TW14}
F_{1}(t) = F_{2}^{1/2}(t)E(t), \quad F_{4}(t) = \frac{1}{2}\left\{E(t) + \frac{1}{E(t)}\right\}F_{2}^{1/2}(t),
\end{equation}
where
$$
E(t) = \exp\left\{ - \frac{1}{2}\int_{t}^{\infty}u(s)ds\right\},
$$
and $u(x)$ is the same Hastings-McLeod solution to the second Painlev\'e transcendent{\footnote{A similar  representation
for the sine-kernel determinant (involving, this time,  a special solution of the fifth Painlev\'e equation) was given earlier in \cite{JMMS}.}}.
Painlev\'e equations are integrable in the sense of Lax pairs, which, in particular, implies that their
solutions admit  a  Riemann-Hilbert (RH) representation. A Riemann-Hilbert representation
can be viewed as the non-abelian analog of the familiar integral representations of
the classical special functions such as the Bessel functions, the
Airy function, etc.  A key consequence of the RH representation is  that
the Painlev\'e functions possess  one of the principal features of a  classical special
function---a mechanism, viz. the nonlinear steepest descent method, to evaluate  explicitly relevant asymptotic connection formulae
(see e.g. \cite{FIKN}).
Specifically, in the case of the Hastings-McLeod solution  the integrability
of the second   Painlev\'e equation underlies  the fact that, in addition to
the asymptotic behavior (\ref{painleve2}) at $t = +\infty$, one also knows
the asymptotic behavior of $u(t)$ at $t = -\infty$, which is described \cite{hast_mcleod} as follows{\footnote{
Hastings and McLeod  derived (\ref{painleve3}) using the inverse scattering transform which was a precursor  of the Riemann-Hilbert method.}}
\begin{equation}\label{painleve3}
u(t) = \sqrt{-\frac{t}{2}} + O\Bigl((-t)^{-5/2}\Bigr), \quad t \to -\infty.
\end{equation}
Herein lies  the importance of the Tracy-Widom formulae (\ref{TW1}) and (\ref{TW14})---they provide the key distribution functions of random matrix theory with {\it explicit} representations
that are amenable to detailed asymptotic analysis.

An immediate corollary of formula (\ref{TW1}) (and the known asymptotics (\ref{painleve3})
of the Painlev\'e function) is an explicit formula for the large negative
$t$ behavior of the distribution function $F_{2}(t)$:
\begin{equation}\label{asympairy1}
\ln F_2(t) = \frac{t^3}{12} - \frac{1}{8}\ln(- t )+ c_0 + o(1), \quad t \to -\infty.
\end{equation}
The value of the constant $c_0$ was conjectured by Tracy and Widom  in paper \cite{75}  to
be  the same as for the sine-kernel determinant, which is given by
\begin{equation}\label{Airyc0}
c_0 = \frac{1}{24}\ln 2 + \zeta'(-1),
\end{equation}
where $\zeta(s)$ is the Riemann zeta-function{\footnote{This conjecture was independently
proved in \cite{DIK} and \cite{Baik:2007}. Also, in \cite{Baik:2007} similar asymptotic results
were established for the other two Tracy-Widom distributions.}}.

Another important advantage of  the  Tracy-Widom formula
(\ref{TW14}) is that it makes it possible to design, using the exact connection
formula (\ref{painleve2})-(\ref{painleve3}),  a very efficient
scheme \cite{Dieng}, \cite{Dris} for the numerical evaluation{\footnote{
It should be noted that
the Hastings-McLeod solution is very unstable; indeed, a small
change in the pre-exponential numerical factor in  the
normalization condition (\ref{painleve2}) yields a completely
different type of  behavior of $u(x)$ for negative $x$, including
the appearance of singularities (e.g. see again \cite{FIKN} ). Herein lies the
importance of the knowledge of the behavior (\ref{painleve3})
in order to adjust  the numerical procedure appropriately.
Said differently,  knowledge of the connection formulae
makes it possible to transform an unstable ODE initial value problem
into a stable ODE boundary value problem.}} of the distribution
functions $F_{\beta}$.


The appearance of the Painlev\'e equations in the Tracy-Widom formulae
is not accidental.  As  follows from  results in  \cite{80},
the   connection to  integrable systems is already
encoded in the determinant formula
(\ref{Airy2}). Indeed, the Airy-kernel integral operator belongs
to a  class of integral operators with kernels of the form
\begin{equation}\label{intker0}
K(z,z') = \frac{\phi(z)\psi(z') - \psi(z)\phi(z') }{z-z'},
\end{equation}
acting in $L_{2}(J)$, where $J$ is a union of intervals,
$$
J = \bigcup_{k=1}^{l}[a_{2k-1}, a_{2k}],
$$
and $\phi(z)$, $\psi(z)$ are $C^{1}$-functions. This class of  integral operators
has  appeared frequently in many applications related to random matrices
and statistical mechanics{\footnote{
An integral operator with kernel (\ref{intker0}) is a special case
of a so-called ``integrable Fredholm operator'', i.e. an integral operator whose
kernel is of the form
\begin{equation}\label{intker2}
K(z,z') = \frac{\sum_{j=1}^{m}f_j(z)h_{j}(z')}{z-z'}, \quad z, z' \in C,
\end{equation}
with some  functions $f_{j}(z)$ and $h_{j}(z)$ defined on a contour $C$.
This type of  integral operators  was singled out as a distinguished class
in \cite{iiks} (see also \cite{deift}; in a  different context,  unrelated to integrable
systems,  these operators were also studied  in the earlier work \cite{Sakh}).
 A crucial property of a kernel (\ref{intker}) is that the associated
resolvent kernel is again an integrable kernel. Moreover,
the functions  $``f"$ and $``h"$ corresponding to the resolvent are
determined via  an auxiliary matrix Riemann-Hilbert problem whose
jump-matrix is explicitly constructed in terms of the original
$f, h$-functions.}}.
In \cite{80}, it is shown that if the functions $\phi(z)$ and $\psi(z)$
satisfy a linear differential equation,
$$
\frac{d}{dz}\begin{pmatrix}\phi(z)\\ \psi(z)\end{pmatrix}
=\Omega(z)\begin{pmatrix}\phi(z)\\ \psi(z)\end{pmatrix},
$$
where
$$
\Omega(z) = \frac{1}{m(z)}\begin{pmatrix}A(z)&B(z)\\C(z)&-A(z)\end{pmatrix},
$$
and $m(z)$, $A(z)$, $B(z)$ and $C(z)$ are polynomials, then
the Fredholm determinant, $\det(1 - K)$, can be expressed in
terms of the solution to a certain system of nonlinear partial differential equations with
the end points $a_j$ as independent variables. This system
is integrable in the sense of Lax{\footnote{The Lax-integrability of the Tracy-Widom system associated
with the kernel (\ref{intker0}) was proven in \cite{P}, where this system was
identified as a special case of   isomonodromy deformation equations
and the Fredholm determinant $\det(1-K)$ was identified  as the corresponding $\tau$-function.}},
and in the case of the Airy kernel it reduces to a single ODE---the second Painlev\'e equation.

In \cite{82}, the results of \cite{80} were extended  to the general $m\times m$ case,
that is, to the kernels of the form
\begin{equation}\label{intker}
K(z,z') = \frac{\sum_{j,i=1}^{m}c_{ij}\phi_j(z)\phi_{i}(z')}{z-z'},
\end{equation}
acting in $L_{2}(J)$. Here,  $J$ is, as before,  a union of intervals,
the constant matrix $ C \equiv \{c_{ij}\}$ is antisymmetric and $\phi_j(z)$ are $C^{1}$-functions
satisfying the linear differential equation
\begin{equation*}
m(z)\frac{d\phi(z)}{dz} = \Omega(z)\phi(z), \quad
\phi(z) := \begin{pmatrix}\phi_1(z)\\\phi_2(z)\\.\\.\\.\\ \phi_m(z)\end{pmatrix},
\end{equation*}
where $m(z)$ is a scalar polynomial while $\Omega(z)$ is
a matrix  with polynomial entries connected to the matrix $C$
by
$$
\Omega^{T}(z)C + C\Omega(z) =0.
$$
The main result of \cite{82} is the derivation, for such  operators $K$,  of a system of
partial differential equations, with the $a_j$ as independent variables,
whose solution  determines the logarithmic
derivatives  of  $\det (1-K)$ with respect to $a_j$.

\begin{figure}[ht]
\centering
\scalebox{0.2}{\includegraphics{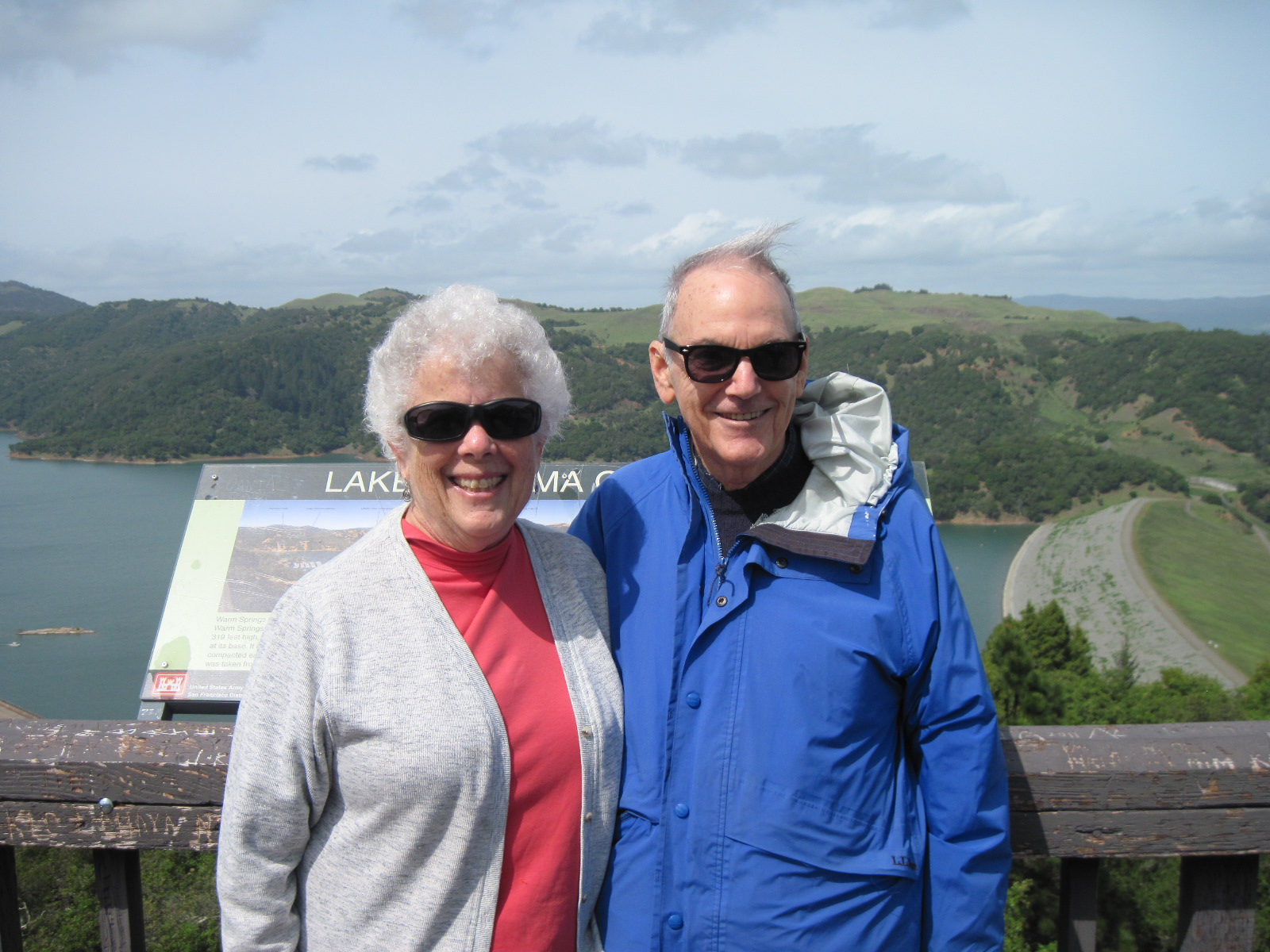}}
 \captionsetup{width=.9\linewidth}
\caption{Harold and his wife, Linda Larkin, at Lake Sonoma in 2013. Photo courtesy of Craig Tracy.}
\label{fig:sine}
\end{figure}

\section{Universality of the Tracy-Widom Distributions}
A remarkable fact is that the Tracy-Widom distribution functions
$F_{\beta}(z)$ appear in a large and  growing number of applications in
random matrix theory as well as in areas beyond
 random matrices.  In this section we will describe
some of these applications.

\subsection{General invariant ensembles and the  Wigner ensembles}
The GUE, GOE and GSE ensembles are respectively particular examples of   unitary  (UE),
orthogonal (OE) and symplectic (SE) ensembles   of random matrix theory.
The joint probability density
of the eigenvalues for these ensembles are given by  (cf. (\ref{eigenprob}))
\begin{equation}\label{eigenprob1}
P_{N \beta}(\lambda_1, ..., \lambda_N) =
Z^{-1}_{N\beta}\,e^{-\eta_{\beta}\sum_{j=1}^{N}V(\lambda_j)}
\prod_{j<k}|\lambda_j - \lambda_k|^{\beta},
\end{equation}
where $\eta_{1}=1$, $\eta_{2} = 1$, $\eta_4 = 2$ (this reflects the fact that the eigenvalues of self-dual Hermitian
matrices come in pairs), and the potential $V(\lambda)$ is a polynomial
with even leading term,
\begin{equation}\label{vgen}
V(\lambda) = \sum_{k=1}^{2m}t_k\lambda^k, \quad t_{2m} > 0.
\end{equation}
(More general potential functions are also allowed.) GUE, GOE and GSE correspond
to the choice $V(\lambda) = \lambda^2$.  Consider again
the upper edge fluctuations of the  spectrum.  {\it Edge-universality}
states that
for a given polynomial potential $V(\lambda)$ there exist scalars   $z_N^{(\beta)}$ and $s_{N}^{(\beta)}$ such that
\begin{equation}\label{betaedge}
\lim_{N\to \infty}\textrm{Prob}\,\left(\frac{\lambda_{\textrm{max}} -z^{(\beta)}_N}{s^{(N)}_{N}}
\leq t\right) = F_{\beta}(t),
\end{equation}
where $F_{\beta}(t)$, $\beta = 1, 2, 4$ are  {\it exactly} the same Tracy-Widom distribution functions
as in (\ref{Airy1}). For $\beta = 2$ this result was obtained  in \cite{DKMVZ1,DKMVZ2} and for $\beta= 1, 4$
in  \cite{DeiGio1,DeiGio2}. It should be noted that the authors  of \cite{DeiGio1,DeiGio2}
based their analysis on Widom's papers \cite{95,96} where  important relations between
symplectic, orthogonal and unitary ensembles were established for finite $N$. These
relations allowed the authors of   \cite{DeiGio1,DeiGio2} to use, in the
symplectic and orthogonal cases, the
asymptotic estimates previously obtained in \cite{DKMVZ2} for the unitary ensembles.
There are also a variety of universality results for eigenvalues in the bulk of the spectrum
(see e.g.  \cite{deift2} and references therein ) for UE, OE, and SE.

Complementary to the invariant ensembles (UE, OE, and SE)  are the so-called  {\it Wigner matrix ensembles}, i.e.,
random matrices with independent identically distributed entries. Tracy-Widom edge universality for these
ensembles is also valid and this important fact  was proved by Soshnikov \cite{Sos}.
In addition, universality in the bulk of the spectrum  for the Wigner ensembles
is now established---see e.g. \cite{Erd} for a comprehensive survey of the results and a historical review.

The Tracy-Widom distribution appears also in  {\it Hermitian matrix models with varying weights}.
These are   unitary ensembles  with  the potential $V(\lambda)$
in (\ref{eigenprob1}) replaced by $NV(\lambda)$. Universality results of the
form (\ref{betaedge}) are also true for such scaled potentials
(see e.g. \cite{DKMVZ1}, \cite{deift2} and the references therein).
For potentials $NV(\lambda)$ a key issue  is
the large $N$ asymptotics of the partition functions
\begin{equation}\label{part0}
Z_N = \int_{-\infty}^{\infty}...\int_{-\infty}^{\infty}e^{-N\sum_{j=1}^{N}V(\lambda_j)}
\prod_{j<k}(\lambda_j - \lambda_k)^{2}d\lambda_1...d\lambda_N.
\end{equation}
Under certain regularity conditions, the leading large $N$  behavior of
the partition function is described by the relation \cite{Joh}
\begin{equation}\label{part1}
Z_N \sim e^{-N^2F_0},\quad N \to \infty,
\end{equation}
where $F_0$ is the {\it free energy} of the model, and  is
given explicitly in terms of the
equilibrium measure corresponding to the potential $V(\lambda)$.
The equilibrium measure is the unique probability measure $\nu_{\textrm{eq}}$
which minimizes the energy functional
$$
I_{V}(\nu) = -\int\int_{\mathbb{R}^2}\ln|\lambda-\lambda'|d\nu(\lambda)d\nu(\lambda')
+ \int_{\mathbb{R}}V(\lambda)d\lambda.
$$
The formula for the free energy reads
$$
F_0 = I_{V}(\nu_{\textrm{eq}}).
$$
As shown in \cite{DKM}, in the case that $V(\lambda)$ is real analytic and
$\frac{V(\lambda)}{\ln |\lambda|} \to + \infty$ as  $|\lambda| \to \infty$,
 the equilibrium measure is absolutely continuous and is supported on a finite
number of intervals.  The derivative of $\nu_{\textrm{eq}}$   coincides with  the mean limiting density of
eigenvalues of the associated UE. If the support of the equilibrium measure consists
of one interval, then one can prove (see  \cite{EM} and also \cite{BI}) the existence
of a full asymptotic expansion,
\begin{equation}\label{part2}
\ln \frac{Z_{N}}{Z_{N}^{\textrm{Gauss}}} \sim -N^2e_0 + \sum_{g=1}^{\infty}\frac{e_{g}}{N^{2g-2}}, \quad N \to \infty,
\end{equation}
where $Z_{N}^{\textrm{Gauss}}$ stands for the partition function of  GUE
( $V(\lambda) = \lambda^2$). Since 1980, the
expansion (\ref{part2}) has played  a  prominent role in the application
of random matrices to enumerative topology. This role was first recognized
in \cite{BIZ}  together with the discovery of its deep connections to the counting
of graphs on  Riemann surfaces  (see \cite{EM} for more details).

The number of  intervals in the support of the  equilibrium measure
depends on the values of the parameters $t_1, ..., t_{2m}$
in (\ref{vgen}). It is of great interest to
study the transition regimes, i.e. regimes involving values of $t_j$
near  critical values  when the number
of the intervals in the support of the equilibrium measure changes.
The first example where one can observe this important  critical
phenomenon is the  even quartic potential $V(\lambda) =
t_4\lambda^4 + t_2\lambda^2$. Up to a trivial renormalization, the
potential $V$  can
be chosen in the form
\begin{equation}\label{quartic1}
V(\lambda) = \frac{1}{4t^2}\lambda^4 + \left(1-\frac{2}{t}\right)\lambda^2.
\end{equation}
For $t>1$ the support of the  equilibrium measure corresponding  to
this potential consists of one interval.
Moreover, as shown in \cite{BI}, all the coefficients $e_{g}(t)$ of the
asymptotic series (\ref{part2}), as functions of $t$,  are real analytic on $[1, \infty)$.
If $t <1$, the support becomes   two intervals.
At the critical value, $t = 1$, the support consists of one interval but the
density function has a double zero inside the support at $\lambda =0$.
The relevant double scaling limit near the critical point $t =1$ is prescribed
by the scaling condition
\begin{equation}\label{scail}
\Bigl|(t-1)N^{2/3}\Bigr| < C.
\end{equation}
As $N \to \infty$ and  $ t \to 1$ satisfying (\ref{scail}), the partition function  admits the
following asymptotic representation \cite{BI}:
\begin{equation}\label{partas}
\frac{Z_N(t)}{Z^{\textrm{Gauss}}} = F_2\Bigl((t-1)2^{2/3}N^{2/3}\Bigr)Z^{\textrm{reg}}_{N}(t)\Bigl(1 + O(N^{-1/3 + \epsilon})\Bigr),\quad \epsilon >0
\end{equation}
where $\ln Z^{\textrm{reg}}_{N}$ is the sum of the first two terms of the series (\ref{part2}), i.e.,
$$
Z^{\textrm{reg}}_{N}(t) = e^{-N^2e_0(t) + e_1(t)},
$$
and $F_2(x)$ denotes, as before, the second Tracy-Widom distribution. Thus we see that in this context $F_2$ appears not as a distribution function but rather as a special factor describing the transition between the one interval and two interval asymptotic regime in the large $N$ limit of these partition functions. This gives $F_2$ added meaning in the realm of enumerative topology, not just probability.
Although not yet proven, it is expected that equation (\ref{partas}) is universal.

\subsection{Random permutations}
Let $\pi = (\pi(1), \pi(2), ..., \pi(N)) $ be a permutation of the numbers
$1, 2, ..., N$. If $1 \leq i_1 <  i_2 < ... < i_k \leq N$ and
$\pi( i_1) < \pi( i_2) < ... < \pi(i_k)$, we say that $\pi( i_1), \pi( i_2), ...,\pi(i_k)$
is an {\it increasing subsequence } in $\pi$ of length $k$. Denote by $l_{N}(\pi)$
the maximal length of all of the increasing subsequences in $\pi$. Suppose that the
permutations are random and uniformly distributed. The principal interest is  in the
limiting statistics of the random variable $l_{N}(\pi)$. This is a subject with a long history
(see e.g. \cite{AD}) which started in the early 60's  with ``Ulam's problem'': prove that
the following limit exists
\begin{equation}\label{ulam}
\lim_{N\to \infty}\frac{{\Bbb E}(l_N)}{\sqrt{N}} = c,
\end{equation}
where ${\Bbb E (\cdot)}$ means mathematical expectation,
and compute $c$.
The proof  of this result  was obtained independently  in  \cite{VK}
and \cite{logan}: It turns out that $c=2$.  Further progress was obtained only  after 22 years in
\cite{BDJ}, where it was shown  that the random variable
$\chi_{N} := \frac{l_N - 2\sqrt{N}}{N^{1/6}}$
converges   in distribution to the {\it second Tracy-Widom  distribution  $F_2(x)$}, i.e.
\begin{equation}\label{bdj}
\lim_{N \to \infty}\textrm{Prob}\left(\frac{l_N - 2\sqrt{N}}{N^{1/6}} \leq t\right) = F_2(t),
\quad \forall t \in {\Bbb R}.
\end{equation}
In addition, the authors of \cite{BDJ} proved convergence of moments:
\begin{equation}\label{bdj2}
\lim_{N\to \infty}{\Bbb E}\Bigl(\chi_{N}^{m}\Bigr) = {\Bbb E}\Bigl(\chi^{m}\Bigr),
\quad \forall m = 1, 2, ...\,.
\end{equation}

It is worth noting that Harold Widom suggested  in \cite{wmoments} an alternative proof of the
convergence of the moments  (\ref{bdj2}) based on the Borodin-Okounkov Fredholm determinant
formula for Toeplitz determinants. The latter was first derived in  \cite{BoOk}, and has been proven extremely
useful in the theory and application  of Toeplitz determinants. There are now several different  proofs  of this  formula and the first
one (after the original proof of Borodin and Okounkov)  was again suggested by Harold Widom jointly with Estelle Basor in
\cite{wb}.  It is remarkable that this proof in turn is again based on one of the old papers of Widom \cite{wbo}. We also refer to the article by Basor, B\"ottcher and Ehrhardt \cite{BBE} in this volume for more
details on the Borodin-Okounkov formula and its history and connection with Widom's work on Toeplitz determinants.
It is important to note that the Borodin-Okounkov formula was in fact discovered
earlier by Geronimo and Case in \cite{GC}, but its significance was overlooked at the time.

Random permutations appear in an enormous array of problems.
Some of the most  fundamental connections are with measures on Young diagrams, random growth processes, interacting particle systems, last passage percolation models and with various tiling problems.
Formula (\ref{bdj}) and its generalizations for more complex types
of permutations,  which involve $F_{\beta}$ with $\beta = 1$ and $4$ as well (see \cite{BR}),  paved the way
that brought the Tracy-Widom distribution functions into many areas of physics, mathematics
and engineering.
We refer to the surveys \cite{deift2,CorwinBAMS} for a discussion of a wide variety of examples and a historical review.


We close this section by mentioning one remarkable generalization of the Ulam problem. Consider $N$-independent standard one-dimensional Brownian motions $B_1(t),B_2(t),\ldots, B_N(t)$, for $t\geq 0$. For $0\leq a<b$ let $B_i(a,b) = B_i(b)-B_i(a)$ denote the increment of the $i^{th}$ Brownian motion on the time interval $[a,b]$. Define the last passage time $M^N_t$ through the environment of Brownian motions by
$$
M^N_t := \max_{0\leq s_1\leq \cdots\leq s_{N-1}\leq t} \left(B_1(0,s_1)+B_2(s_1,s_2)+\cdots + B_{N}(s_{N-1},t)\right).
$$
In joint work \cite{105} with Janko Gravner and Craig Tracy, Harold Widom proved that $M^N_t/\sqrt{t}$  has the same distribution as the largest eigenvalue of an $N\times N$ GUE matrix. In fact, in their work, they also study some discrete generalizations of the Ulam problem which are related to growth models and last passage problems such as above.

\section{The asymmetric simple exclusion process. Harold Widom's papers  \cite{133,134,135}}
The asymmetric simple exclusion process (ASEP) was introduced (in the probability literature) in Spitzer's 1970 work \cite{Sp}. It is a continuous time Markov process of interacting particles on the integer lattice ${\Bbb Z}$. A particle at a location $x$ waits an exponential time
with parameter one (independently of all other particles). When that time has elapsed, the particle flips a coin and with probability $p$ attempts to jump one position to the right, and with probability $q=1-p$ attempts to jump one position to the left. These jumps are achieved only if the destination site is unoccupied at that time. Due to the memoryless property of the exponential distribution, this defines a Markov process. See Figure \ref{fig:asep} for an illustration of ASEP. Since its appearance, the ASEP model has  attracted immense attention in both the mathematical and physics communities, due to the fact that this is one of the simplest nontrivial processes modeling nonequilibrium phenomena.

\begin{figure}[h]
\centering
\scalebox{0.6}{\includegraphics{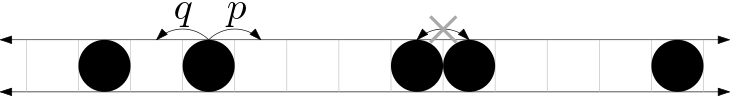}}
 \captionsetup{width=.9\linewidth}
\caption{In the ASEP, particles jump left and right according to exponential clocks with rates $q$ and $p$. All jumps are independent, and those that would lead to collisions are excluded (the grey $\times$).}
\label{fig:asep}
\end{figure}

When $p=0$, the model is known as TASEP, with the $T$ standing for ``totally'' since now particles only march to the left.
In the   paper \cite{Joh2}, building on  \cite{BDJ}, an important relation of the particle probabilities of the totally asymmetric process
(TASEP, $p=0$) to the distribution function of the largest eigenvalue in the Laguerre unitary random matrix ensemble was discovered. Relying on tools from random matrix theory (namely, orthogonal polynomials and Fredholm determinants) \cite{Joh2} shows that the limiting particle distribution in  TASEP is given by the Tracy-Widom $\beta=2$ distribution (see also \cite{PS}).
 The decade after \cite{BDJ} saw incredible developments in the analysis of models like TASEP and the longest increasing subsequence whose analysis reduces to studying asymptotics of determinants (these models are called ``determinantal'', see \cite{BorodinDet}).

The extension of this result to ASEP when $q\neq p>0$ posed an enormous challenge since ASEP,
as opposed  to TASEP, is not determinantal and a direct relationship to random matrices is no longer available.
The extraordinary series of papers  \cite{133,134,135},
nevertheless  provides this extension by building on the relationship between ASEP and
the Heisenberg XXZ spin chain and exploiting the ideas of the coordinate Bethe Ansatz\footnote{This method dates back to 1931 work of Hans Bethe \cite{Bethe} in studying the Heisenberg XXX spin chain \cite{Heis}.}.
These works broke new ground as the first instance of a non-determinantal model for which the Tracy-Widom distributions was demonstrated.


The starting point for Tracy and Widom's work on ASEP is an exact formula for the transition probability of the $N$-particle ASEP. For $N\in {\Bbb N}$, consider a system of $N$ particles on ${\Bbb Z}$. For two vectors of ordered integers, $\vec{x} = \{x_1<\cdots< x_N\}$, and  $\vec{y} = \{ y_1<\cdots< y_N\}$, let $P_{\vec{y}}(\vec{x};t)$ denote the probability that ASEP starts in state $\vec{y}$ at time zero, and is in state $\vec{x}$ at time $t$. This function solves the ``Master equation'' (or Kolmogorov forward equation). When $N=1$, this equation reads
$$
\frac{d}{dt}P_{y}(x;t) = \Delta_{p,q} P_{y}(x;t),\qquad P_{y}(x;0)=\mathbf{1}_{x=y}
$$
where $\Delta_{p,q}$ acts in the $x$ variable as $\Delta_{p,q}f(x) = pf(x-1) + qf(x+1)-f(x)$. This equation can be solved explicitly using the Fourier transform, yielding
$$
P_{y}(x;t) = \frac{1}{2\pi i} \int \xi^{x-y-1} e^{\epsilon(\xi)t} d\xi,\qquad \epsilon(\xi) = p\xi^{-1} +q \xi -1
$$
where $\xi$ is integrated along the complex unit circle centered at $0$.

When $N\geq 2$, the Master equation becomes more complicated. When the $x_i$ are well-spaced (at least distance two apart), the right-hand side becomes the sum of $N$ terms, each involving $\Delta_{p,q}$ acting on each of the $x_i$ variables. However, when the $x_i$ neighbor each other, certain terms need to be excluded (corresponding to the excluded jumps in ASEP). Thus, the system becomes non-constant coefficient and non-separable and so there is no a priori reason to expect such a simple contour integral formula as above.

There is hope, however. The generator of ASEP (the operator on the right-hand side of the Master equation) is related by a similarity transform to the Hamiltonian of the XXZ quantum spin chain, itself a special limit of the six-vertex model. These models have a long history in the realm of exactly solvable models in statistical mechanics, e.g. \cite{Baxter}, and it was known since \cite{Lieb} that the coordinate Bethe ansatz can be used to diagonalize the six-vertex model transfer matrix. For the six-vertex model and XXZ spin chain, one has to assume boundary conditions (e.g. work on a torus). However, since the generator of ASEP is stochastic, one can work directly on the full space ${\Bbb Z}$. This actually leads to a significant simplification to the Bethe ansatz---there are no longer complicated Bethe equations! Instead, one has to develop an analog of the classical Plancherel theory, now for the Bethe ansatz eigenfunctions. This is essentially accomplished by \cite{133} as well as \cite{BCPS}.

\begin{figure}[h]
\centering
\scalebox{0.16}{\includegraphics{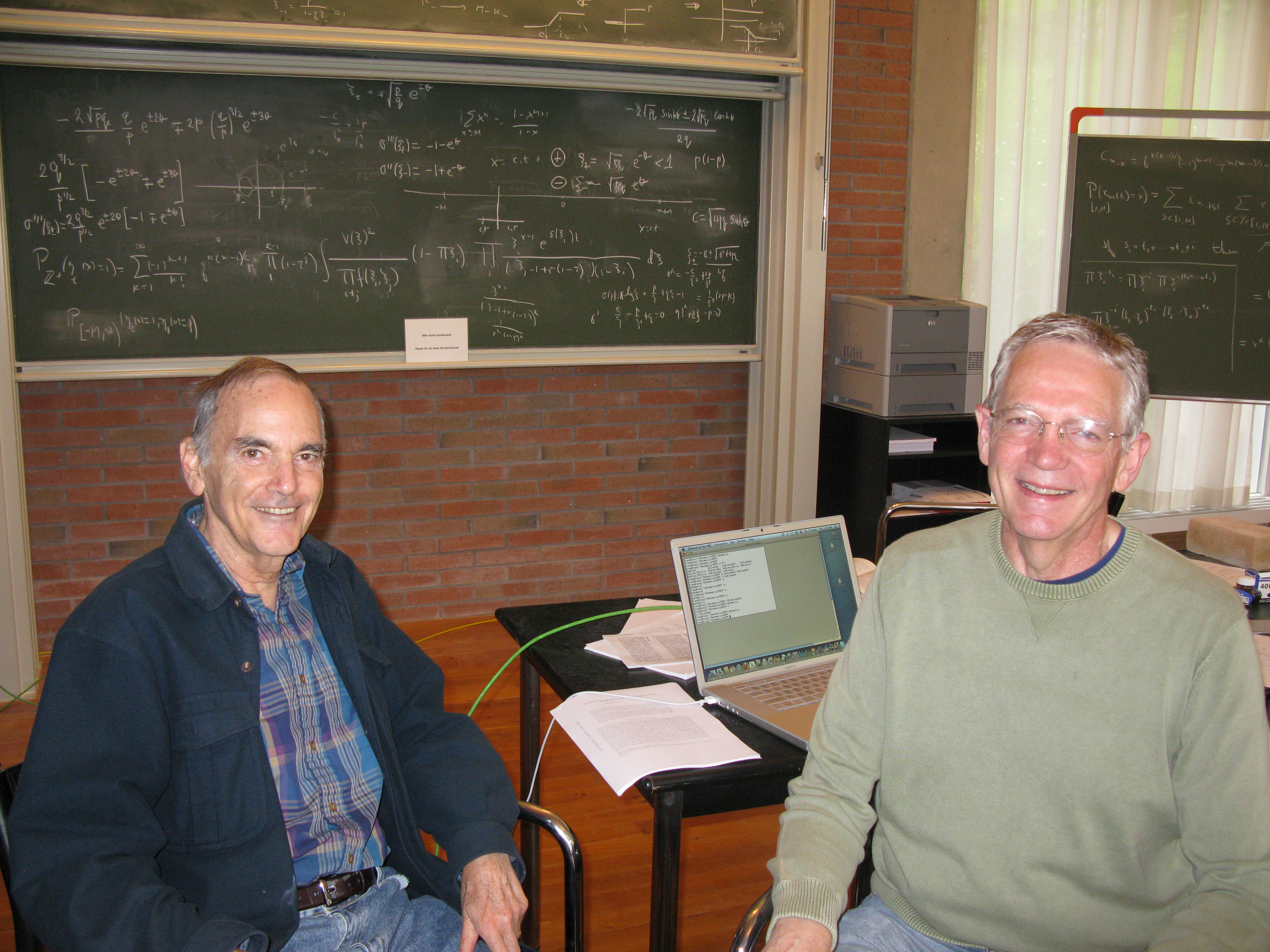}}
 \captionsetup{width=.9\linewidth}
\caption{Harold Widom with his long-time collaborator, Craig Tracy at Oberwolfach in May 2009. Seen on the blackboard is calculations related to their work on ASEP. Photo courtesy of the Archives of the Mathematisches Forschungsinstitut Oberwolfach.}
\label{fig:asep}
\end{figure}

The following formula was proved in \cite{133}. The $N=2$ case had previously appeared in work of R\'akos and Sch\"utz \cite{RakosSchutz}. Provided $p\neq 0$,
\begin{equation}\label{eq:tranprob}
P_{\vec{y}}(\vec{x};t) = \sum_{\sigma \in S_N} \frac{1}{(2\pi i)^N}\int\cdots \int A_{\sigma} \prod_{j=1}^{N} \xi_{\sigma(j)}^{x_j-y_{\sigma(j)}-1} e^{\epsilon(\xi_j)t} d\xi_j.
\end{equation}
Here, the sum $\sigma\in S_N$ is taken over all permutations $\sigma$ on $N$ elements; the  $\xi_j$ variables are complex and integrated over contours which are assumed to be circles around the origin with radii small enough so as not to contain any poles of the $A_\sigma$ term; this $A_{\sigma}$ term is given by the formula
$$
A_{\sigma} = \prod \left\{S_{\alpha\beta}:\{\alpha,\beta\}\textrm{ is an inversion of }\sigma\right\},\qquad S_{\alpha\beta} = - \frac{p+q \xi_{\alpha} \xi_{\beta} - \xi_{\alpha}}{p+q \xi_{\alpha} \xi_{\beta} - \xi_{\beta}}.
$$
where the product is of terms $S_{\alpha\beta}$ over all pairs $1\leq \alpha<\beta\leq N$ such that $\sigma(\alpha)>\sigma(\beta)$.

This sum over the symmetric group may seem reminiscent of the expansion formula for determinants. In the special case when $p=1$ (which is the same at taking $p=0$, up to some symmetry), there is factorization and $A_\sigma = (-1)^{\sigma} \prod_{j=1}^{N} (1-\xi_{\sigma(j)})^{\sigma(j)-j}$. This implies that for $p=1$,
$$
P_{\vec{y}}(\vec{x};t)=\det\left(\frac{1}{2\pi i} \int (1-\xi)^{j-i} \xi^{x_i-y_j-1}e^{\epsilon(\xi)t} d\xi\right)_{i,j=1}^{N}
$$
which recovers one aspect of the determinantal structure for TASEP, see \cite{Schutz}.

Returning to (\ref{eq:tranprob}), just like in random matrix eigenvalue measures, the challenge now becomes how to extract large $N$ asymptotics for marginals of this measure on point configurations. Random matrix methods, though informative to this aim, were not directly applicable. The paper \cite{133} extracted summation formulas for the marginal distribution of $x_m$ for general $N$ and initial data $\vec{y}$. Taking $y_i=i$ and then taking $N$ to infinity led Tracy and Widom to study ``step'' initial condition for ASEP in which initially every positive integer site is occupied. The paper \cite{133} closed with an infinite summation formula for the marginal distribution of the location of the $m^{th}$ left-most particle at time $t$, for any $m$. This formula was further manipulated in \cite{134} into an integral transform of a Fredholm determinant. They had discovered a hidden determinantal structure in this model---one which would then allow them to extract asymptotics in \cite{135}. It should be remarked that the analysis of the Fredholm determinant in \cite{135} was itself quite novel, requiring a number of ingenious operator manipulations.

Let us now record the main result of the asymptotics performed in \cite{135}. In fact, in \cite{139}, a more general class of initial data was considered in which sites in ${\Bbb N}$ are inititally occupied according to independent Bernoulli coin flips with probability $\rho\in (0,1]$ of having a particle. The case $\rho=1$ recovers the step initial data.

Assume that $0\leq p < q$ and put $\gamma := q-p$. Denote also $\sigma =  m/t \leq 1$ and
define the constants,
$$
c_1 := -1 + 2\sqrt{\sigma}, \quad c_2 := \sigma^{-1/6}\Bigl(1 - \sqrt{\sigma}\Bigr)^{2/3}.
$$
Then \cite{135,139} shows that (cf.  (\ref{betaedge}) and (\ref{bdj})),
\begin{equation}\label{asep1}
\lim_{t \to \infty}\textrm{Prob}\left(\frac{x_m\Bigl(t/\gamma\Bigr) - c_1t}{c_2t^{1/3}} \leq s\right)
= F_2(s), \quad \textrm{when}\quad 0 < \sigma < \rho^2,
\end{equation}
and
\begin{equation}\label{asep2}
\lim_{t \to \infty}\textrm{Prob}\left(\frac{x_m\Bigl(t/\gamma\Bigr) - c_1t}{c_2t^{1/3}} \leq s\right)
= F_1(s)^2, \quad \textrm{when}\quad \sigma = \rho^2, \quad \rho < 1.
\end{equation}

We emphasize again that ASEP is not a determinantal process and hence none of
the usual ``integrable techniques'', such as the orthogonal
polynomial approach, the  Riemann-Hilbert method, the Borodin-Okounkov
formula, or the  theory of  integrable
Fredholm operators are readily applicable. In spite of the absence of such
tools,  the Tracy-Widom distributions still arise!

Besides demonstrating that the Tracy-Widom distribution is universal for ASEP regardless of the choice of parameters (provided $p\neq q$), the work in \cite{133,134,135} demonstrated that non-determinantal models could still be solved, albeit with different techniques.

Over the decade or so since Tracy and Widom's initial work on ASEP, there have been a number of developments in the field of integrable probability in this direction. This is too vast a subject to try to survey here, so we simply refer to a few existing surveys \cite{BorodinWheelerAggarwal, BorodinICM, BorodinGorin, BorodinPetrov1,BorodinPetrov2, BorodinWheeler, CorwinKPZ,  CorwinTwoASEP, CorwinICM,CorwinBAMS,OConnell,  QuastelSpohn, Zygouras1812.07204} and mention some general topics: Kardar-Parisi-Zhang universality class, replica method, Markov duality, quantum Toda Hamiltonian, Macdonald processes (and limits to Whittaker, Jack, Hall-Littlewood or Schur processes), stochastic vertex models, spin $q$-Whittaker and Hall-Littlewood functions. The list goes on, as does the influence of Tracy and Widom's work on ASEP and random matrices.





\begin{thebibliography}{2}
\centerline{\Large{Cited Publications by Harold Widom}}
\vskip .2in
\bibitem{33}  {\it The strong Szeg\H{o} limit theorem for circular arcs.} Indiana Univ. Math. J.  21,
277--283 (1971).
\bibitem{wbo}{\it Toeplitz determinants with singular generating functions.} Amer. J. Math. 95,
 333--383 (1973).
\bibitem{72}{\it The asymptotics of a continuous analogue of orthogonal polynomials}. J. Approx. Th.
76, 51--64 (1994).
\bibitem{105}(with J. Gravner and C. A. Tracy) {\it Limit theorems for height fluctuations in a class of discrete space and time growth models}. J. Stat. Phys. 102, 1085--1132 (2001).
\bibitem{74}(with C. A. Tracy) {\it Level-spacing distributions and the Airy kernel}. Phys. Letts. B 305,
 115--118 (1993).
\bibitem{75}(with C. A. Tracy) {\it Level-spacing distributions and the Airy kernel.} Comm. Math.
Phys. 159,  151--174 (1994).
\bibitem{80}(with C. A. Tracy) {\it Fredholm determinants, differential equations and matrix models.}
Comm. Math. Phys. 163, 38--72 (1994).
\bibitem{82}(with C. A. Tracy) {\it Systems of partial differential equations for a class of operator determinants.} Oper. Th.: Adv. Appl. 78, 381--388 (1995).
\bibitem{83}{\it Asymptotics for the Fredholm determinant of the sine kernel on a union of intervals.}
Comm. Math. Phys. 171, 159--180 (1995) .
\bibitem{84}(with C. A. Tracy) {\it On orthogonal and symplectic matrix ensembles}. Comm. Math.
Phys. 177,  727--754 (1996).
\bibitem{95}(with C. A. Tracy){\it  Correlation functions, cluster functions and spacing distributions
for random matrices}.,J. Stat. Phys. 92,  809--835 (1998).
\bibitem{96}{\it On the relation between orthogonal, symplectic and unitary matrix ensembles}. J. Stat.
Phys. 94,  347--364 (1999).
\bibitem{wb}{\it On a Toeplitz determinant identity of Borodin and Okounkov}. Int.
Eqs. Oper. Th. 37, 397-401 (2000).
\bibitem{wmoments} {\it A note on convergence of moments for random Young tableaux and a random growth
model.} Intl. Math. Research Notices. 9,  455--464 (2002).
\bibitem{133}(with C. A. Tracy){\it Integral Formulas for the Asymmetric Simple Exclusion Process.}
Comm. Math. Phys. 279,  815--844 (2008). Erratum: Comm. Math. Phys. 304, 875--878 (2011).
\bibitem{134}(with C. A. Tracy) {\it A Fredholm Determinant Representation in ASEP}. J. Stat. Phys.
132,   291--300 (2008).
\bibitem{135}(with C. A. Tracy) {\it Asymptotics in ASEP with step initial condition}. Comm. Math.
Phys. 290, 129--154  (2009).
\bibitem{139}(with C. A. Tracy) {\it On ASEP with step Bernoulli initial condition.} J. Stat. Phys. 137,
825--838 (2009).
\vskip .2in
\centerline{\Large{ Other References} }
\vskip .2in
\bibitem[ABW]{BorodinWheelerAggarwal} A. Aggarwal, A. Borodin, M. Wheeler, {\it Colored Fermionic Vertex Models and Symmetric Functions}, arXiv:2101.01605 (2021).

\bibitem[AD]{AD} D. Aldous,  P. Diaconis, {\it Longest Increasing Subsequences: From Patience Sorting to the Baik-Deift-Johansson Theorem},   Bull. Amer. Math. Soc. 36 no. 4,  413--432 (1999).

\bibitem[BBD]{Baik:2007}
J. Baik, R.  Buckingham,  and J. DiFranco
{\it Asymptotics of Tracy-Widom distributions and the total integral of a Painlev\'e II function},
Commun. Math. Phys.  280,   463--497  (2008)

\bibitem[BDJ]{BDJ} J. Baik, P.  Deift,  K. Johansson, {\it On the distribution of the length of the longest increasing
subsequence of random permutations},  J. Amer. Math. Soc.  12, 1119--1178  (1999).


\bibitem[BaRa]{BR} J.  Baik, E. M.  Rains, {\it The asymptotics of monotone subsequences of involutions},  Duke
Math. J. 109, 205--281 (2001) .

\bibitem[BBE]{BBE} E. Basor, A. B\"ottcher, and T. Ehrhardt, {\it Harold Widom's work in Toeplitz determinants}.
This issue.


\bibitem[Bax]{Baxter} R. Baxter, {\it Exactly solved models in statistical mechanics}, Dover, 2007.


\bibitem[BIZ]{BIZ} D. Bessis, C. Itzykson, J. B. Zuber, {\it Quantum field theory techniques in
graphical enumeration}, Adv. in Appl. Math., 1 - 2 , 109 -- 157 (1980).

\bibitem[Bet]{Bethe} H. Bethe, {\it Zur Theorie der Metalle. I. Eigenwerte und Eigenfunktionen der linearen Atomkette},
Z. Phys. 71, 205 (1931).

\bibitem[BI]{BI} P. M. Bleher, A. R. Its, {\it Asymptotics of the partition function of
a random matrix model}, Ann. Inst. Fourier, Grenoble  55, 6, 1943--2000 (2005).

\bibitem[BoOk]{BoOk} A. Borodin and A. Okounkov, {\it  A Fredholm determinant formula for Toeplitz determinants.}
Integral Equations Operator Theory 37, 386--396 (2000).

\bibitem[Bor1]{BorodinDet}
A. Borodin, {\it Determinantal point processes}, In: Oxford handbook of random matrix theory. Oxford University Press, 231--249. (2011).

\bibitem[Bor2]{BorodinICM} A. Borodin, {\it Integrable probability}, Proceedings of the 2014 ICM.
\bibitem[BCPS]{BCPS} A. Borodin, I. Corwin, L. Petrov, T. Sasamoto, {\it Spectral theory for interacting particle systems solvable by coordinate Bethe ansatz}, Comm. Math. Phys. 339, 1167--1245 (2015).
\bibitem[BoGo]{BorodinGorin} A. Borodin, V. Gorin, {\it Lectures on integrable probability}, arXiv: 1212.3351, 2012.
\bibitem[BoPe1]{BorodinPetrov1} A. Borodin, L. Petrov, {\it Integrable probability: From representation theory to Macdonald processes}, Probab. Surveys 11, 1--58 (2014).
\bibitem[BoPe2]{BorodinPetrov2} A. Borodin, L. Petrov, {\it Higher spin six vertex model and symmetric rational functions}, Selecta Math. 24, 751--874 (2018).
\bibitem[BoWh]{BorodinWheeler} A. Borodin, M. Wheeler, {\it Coloured stochastic vertex models and their spectral theory}, arXiv:1808.01866, 2018.






\bibitem[Cor1]{CorwinKPZ} I. Corwin, {\it The Kardar-Parisi-Zhang equation and universality class}, Rand. Mat.: Theo. Appl., 1, 113001 (2012).

\bibitem[Cor2]{CorwinTwoASEP} I. Corwin, {\it Two ways to solve ASEP}, In: Topics in Percolative and Disordered Systems, (2014).


\bibitem[Cor3]{CorwinICM} I. Corwin, {\it Macdonald processes, quantum integrable systems and the Kardar-Parisi-Zhang universality class}, Proceedings of the 2014 ICM.

\bibitem[Cor4]{CorwinBAMS} I. Corwin, {\it Commentary on ``Longest increasing subsequences: from patience sorting to the Baik-Deift-Johansson theorem'' by David Aldous and Persi Diaconis}, Bull. Amer. Math. Soc., 55, 363-374 (2018).







\bibitem[Dei1]{Deift0} P. Deift, {\it Orthogonal polynomials and random matrices: a
Riemann-Hilbert approach.} Courant Lecture Notes in Math. 1998

\bibitem[Dei2]{deift} P.~A.~Deift, {\it Integrable operators}, in Differential operators and
spectral theory: M. Sh. Birman's 70th anniversary collection, V. Buslaev,
M. Solomyak, D. Yafaev, eds., American Mathematical Society Translations,
ser. 2 , 189, Providence, RI: AMS, 1999.

\bibitem[Dei3]{deift2} P. Deift, {\it Universality for mathematical and physical systems},
International Congress of Mathematicians,  1, 125 -- 152, Eur. Math. Soc. Z\"urich, 2007.

\bibitem[DeGi1]{DeiGio1} P.  Deift, D.  Gioev, {\it Universality in random matrix theory for orthogonal and symplectic
ensembles.} Inter. Math. Res. Papers, no. 2, Art.IDrpm 004, 116 pp (2007).
\bibitem[DeGi2]{DeiGio2} P.  Deift, D.  Gioev, {\it Universality at the edge of the spectrum for unitary, orthogonal
and symplectic ensembles of random matrices.} Comm. Pure Appl. Math.,  60,  no.6,
867 -- 910  (2007)

\bibitem[Die]{Dieng} M.  Dieng {\it  Distribution Functions for Edge Eigenvalues in Orthogonal and Symplectic Ensembles:
Painlev\'e Representations,}  Ph.D. thesis (2005), University of Davis. arXiv:math/0506586v2.


\bibitem[DIK]{DIK}  P. Deift, A.  Its, I. Krasovsky,
  {\it  Asymptotics of the Airy-kernel determinant,}
 Commun. Math. Phys. 278,  643--678 (2008) .

\bibitem[DKM]{DKM} P. A. Deift, T. Kriecherbauer, K. D. T-R. McLaughlin,
{\it New results on the equilibrium measure for logarithmic potentials in
the presence of an external field}, J. Approx. Theory 95, 388 -- 475 (1998).

\bibitem[DIZ]{DIZ} P. Deift, A. Its, and X. Zhou, {\it A Riemann-Hilbert
  approach to asymptotic problems arising in the theory of random
  matrix models, and also in the theory of integrable statistical
  mechanics.} Ann. Math. 146, 149--235 (1997).

\bibitem[DKMVZ1]{DKMVZ1} P. A. Deift, T. Kriecherbauer, K. D. T-R. McLaughlin, S. Venakides, X.  Zhou,
{\it  Uniform asymptotics for polynomials orthogonal with respect to varying exponential weights and
applications to universality questions in random matrix theory, }  Comm. Pure Appl. Math. 52, 1335 -- 1425 (1999).

\bibitem[DKMVZ2]{DKMVZ2} P. A.  Deift, T.  Kriecherbauer, K. T-R.  McLaughlin, S. Venakides, X. Zhou, {\it Strong
asymptotics of orthogonal polynomials with respect to exponential weights.} Comm. Pure Appl.
Math.  52, 149 -- 1552 (1999).

\bibitem[dCM]{dCM} J. des Cloizeaux and M. L. Mehta,
{\it Asymptotic behavior of spacing distributions for the eigenvalues of
random matrices,} J. Math. Phys., 14,   1648--1650 (1973).



\bibitem[DBT]{Dris} T. A.  Driscoll, F.  Bornemann,  and L. N. Trefethen, {\it The chebop system for automatic
solution of differential equations,}  BIT, 48 , 701--723 (2008).


\bibitem[Dys1]{D1} F. Dyson, {\it Statistical theory of the energy levels of complex systems, parts I, II, III,}
J. Math. Phys. 3,  140 -- 156 (1962).

\bibitem[Dys2]{D2} F. Dyson, {\it Fredholm determinants and inverse scattering problems,}  Commun. Math. Phys.  47,
171-183 (1976).


\bibitem[Erd]{Erd} L. Erdos, {\it Universality of Wigner random matrices: a survey of recent results}
Uspekhi Mat. Nauk 66, Issue 3(399),  67--198  (2011).

\bibitem[Ehr]{E} T. Ehrhardt, {\it Dyson's constant in the asymptotics of the Fredholm
determinant of the sine kernel.} Comm. Math. Phys. 262, 317 -- 341   (2006).

\bibitem[EM]{EM} N. M. Ercolani,  K. D. T-R. McLaughlin, {\it Asymptotics of the partition function for
random matrices via Riemann-Hilbert techniques and applications to graphical enumeration},
Int. Math. Res. Not. 14, 755 -- 820 (2003).

\bibitem[GeCa]{GC} J. S. Geronimo and K. M. Case, {\it Scattering theory and polynomials orthogonal on the unit
circle}, J. Math. Phys. 20, no. 2, 299--310 (1979).

\bibitem[For]{For} P. J. Forrester, {\it The spectrum edge of random matrix ensembles}, Nucl. Phys. B
402, 709 - 728 (1993).


\bibitem[FIKN]{FIKN} A. Fokas, A. Its, A. Kapaev, V. Novokshenov, {\it Painlev\'e Transcendents:
The Riemann-Hilbert Approach,} AMS  Mathematical Surveys and Monographs, vol. 128,
2006.

\bibitem[JMMS]{JMMS}
M. Jimbo, T. Miwa, Y. M\^ori, M. Sato, {\it Density matrix of impenetrable
bose gas and the fifth Painlev\'e transcendent,} Physica D 1, 80--158 (1980).

\bibitem[Hei]{Heis}W. Heisenberg, {\it Zur Theorie des Ferromagnetismus,} Z. Phys., vol. 49, pp. 619-636, 1928

\bibitem[HaMc]{hast_mcleod}
S.~P. Hastings and J.~B. McLeod, \textit{A boundary value problem associated
with the second Painlev\'e transcendent and the Korteweg-de Vries
equation},  Arch.\ rational Mech.\ Anal. 73, 31--51(1980).


\bibitem[IIKS]{iiks} A.~R. Its, A.~ G. Izergin, V.~E. Korepin, N.~ A. Slavnov,
{\it Differential Equations for Quantum Correlation Functions},
J. Mod. Phys. B 4, 1003--1037  (1990).


\bibitem[Joh1]{Joh} K. Johansson, {\it On fluctuations of eigenvalues of random matrices}, Duke
Mathematics Journal 91, no. 1, 151 -- 204  (1998).



\bibitem[Joh2]{Joh2} K.  Johansson {\it  Shape fluctuations and random matrices},  Commun. Math. Phys.
209, 437 -- 476  (2000).

\bibitem[Kra] {K} I. V. Krasovsky, {\it Gap probability in the spectrum of random matrices
and asymptotics of polynomials orthogonal on an arc of the unit
circle.}  Int. Math. Res. Not. 2004 , 1249--1272 (2004).



\bibitem[Len]{L} A. Lenard, {\it One-dimensional impenetrable bosons in termal equilibrium }, J. Math. Phys.
7, 1268 -- 72 (1966).

\bibitem[Lieb]{Lieb} E. H. Lieb, {\it Residual Entropy of Square Ice}, Phys. Rev. 162, 162 (1967).


\bibitem[LoSh]{logan} B. F. Logan and L. A. Shepp, {\it A variational problem for Young tableaux},
Adv. in Math. 26, 206 -- 222  (1977).

\bibitem[Meh]{M} M. L. Mehta, {\it Random matrices}, San Diego, Academic 1990.

\bibitem[OCon]{OConnell} N. O'Connell, {\it Whittaker functions and related stochastic processes}, In: MSRI Volume -- Random Matrix Theory, Interacting Particle Systems and Integrable Systems, vol 65. Ed. P. Deift and P. Forrester.
\bibitem[Zyg]{Zygouras1812.07204} N. Zygouras, {\it Some algebraic structures in the KPZ universality}, arXiv:1812.07204.


\bibitem[Pal]{P} J. Palmer, \textit{Deformation analysis of matrix models}, Physica D 78, 166--185 (1995).

\bibitem[PrSp]{PS} M. Pr\"ahofer, H.  Spohn, {\it  Current fluctuations for the totally asymmetric simple
exclusion process. In and Out of Equilibrium}, Progress in Probability 51, 185 -- 204 (2000).


\bibitem[QuSp]{QuastelSpohn} J. Quastel and H. Spohn, {\it The one-dimensional KPZ equation and its universality class}, J. Stat. Phys. 160, 965--884 (2015).

\bibitem[RaSc]{RakosSchutz} A. R\'akos, G. M. Sch\"utz, {\it Current distribution and random matrix ensembles for
an integrable asymmetric fragmentation process.} J. Stat. Phys. 118, 511--530 (2005).


\bibitem[Sak]{Sakh} L. A. Sakhnovich, {\it Operators similar to unitary operators,} Functional Anal.and Appl. 2, no 1, 48--60 (1968).


\bibitem[Sch]{Schutz} G. M. Sch\"utz, {\it Exact solution of the master equation for the asymmetric exclusion process.}, J. Stat. Phys. 88, 427--445 (1997).

\bibitem[Sos]{Sos} A. Soshnikov, {\it  Universality at the edge of the spectrum in Wigner random matrices}, Comm. Math.
Phys. 207, no.3,  697--733 (1999).


\bibitem[Spi]{Sp} F. Spitzer, {\it  Interaction of Markov processes,}  Adv. Math. 5, 246--290 (1970).

\bibitem[VeKe]{VK} A. M. Vershik and S. V. Kerov, {\it  Asymptotics of the Plancherel measure of the symmetric
group and the limiting form of Young tables}, Soviet Math. Dokl. 18,  527--531 (1977).


\end{thebibliography}
\end{document}